\documentclass[conference]{IEEEtran}
\IEEEoverridecommandlockouts
\usepackage{cite}
\usepackage{amsmath,amssymb,amsfonts}
\usepackage{algorithmic}
\usepackage{graphicx}
\usepackage{textcomp}
\usepackage{xcolor}

\usepackage{fancyhdr}
\usepackage[hyphens]{url}

\usepackage{mathptmx} 
\usepackage[T1]{fontenc}
\newcommand{\ignore}[1]{}
\usepackage[normalem]{ulem}
\usepackage[hyphens]{url}
\usepackage{microtype}
\usepackage{multirow}
\usepackage[skip=3pt]{caption}
\usepackage{float}
\usepackage[hyphens]{url}
\usepackage{mathtools}
\usepackage{xspace}
\usepackage{color}
\usepackage{comment}
\usepackage[flushleft]{threeparttable}
\usepackage{cite}
\usepackage{epstopdf}
\usepackage{pifont}
\usepackage{listings}
\usepackage{tikz}
\usepackage{subdepth}
\usepackage{subcaption}
\usepackage{algorithm}
\usepackage{enumitem}
\setitemize{noitemsep,topsep=0pt,parsep=0pt,partopsep=0pt}
\usepackage{listings}
\usepackage{caption}
\usepackage[hang,flushmargin]{footmisc}

\newcommand{\eg}{\emph{e.g.}\xspace}
\newcommand{\etal}{\textit{et al.}}

\def\authnotes{1}
\newcommand{\authnote}[2]{\ifnum\authnotes=1\begin{quote}\textbf{#1 says:} #2\end{quote}\fi}
\newcommand{\fixme}[1]{\ifnum\authnotes=1\textbf{\textcolor{red}{[FIXME: #1]}}\fi}

\newcommand{\bheading}[1]{{\vspace{2pt}\noindent{\textbf{#1}}\hspace{2pt}}}
\newcommand{\RNum}[1]{\uppercase\expandafter{\romannumeral #1\relax}}%

\usepackage{array}
\newcolumntype{L}[1]{>{\raggedright\let\newline\\\arraybackslash\hspace{0pt}}m{#1}}
\newcolumntype{C}[1]{>{\centering\let\newline\\\arraybackslash\hspace{0pt}}m{#1}}
\newcolumntype{R}[1]{>{\raggedleft\let\newline\\\arraybackslash\hspace{0pt}}m{#1}}%

\newcommand*{\QEDB}{\hfill\ensuremath{\square}}%

\newcommand\blfootnote[1]{%
  \begingroup
  \renewcommand\thefootnote{}\footnote{#1}%
  \addtocounter{footnote}{-1}%
  \endgroup
}

\usepackage[bookmarks=true,breaklinks=true,letterpaper=true,colorlinks,linkcolor=black,citecolor=blue,urlcolor=black]{hyperref}

\def\BibTeX{{\rm B\kern-.05em{\sc i\kern-.025em b}\kern-.08em
    T\kern-.1667em\lower.7ex\hbox{E}\kern-.125emX}}

\begin{document}

\title{New Models for Understanding and Reasoning about Speculative Execution Attacks}

\author{\IEEEauthorblockN{Zecheng He}
\IEEEauthorblockA{\textit{Princeton University} \\
zechengh@princeton.edu}
\and
\IEEEauthorblockN{Guangyuan Hu}
\IEEEauthorblockA{\textit{Princeton University} \\
gh9@princeton.edu}
\and
\IEEEauthorblockN{Ruby Lee}
\IEEEauthorblockA{\textit{Princeton University} \\
rblee@princeton.edu}
}

\maketitle

\begin{abstract}
Spectre and Meltdown attacks and their variants exploit
hardware performance optimization features to cause security breaches. Secret information is accessed and leaked through covert or side channels. New attack variants keep appearing and we do not have a systematic way to capture the critical characteristics of these attacks and evaluate why they succeed or fail.

In this paper, we provide a new attack-graph model for reasoning about speculative execution attacks. We model attacks as ordered dependency graphs, and prove that a race condition between two nodes can occur if there is a missing dependency edge between them. We  define a new concept, ``\textit{security dependency}'', between a resource access and its prior authorization operation. We show that a missing \textit{security dependency} is equivalent to a race condition between authorization and access, which is a root cause of speculative execution attacks. We show detailed examples of how our attack graph models the Spectre and Meltdown attacks, and is generalizable to all the attack variants published so far. This attack model is also very useful for identifying new attacks and for generalizing defense strategies. We identify several defense strategies with different performance-security tradeoffs.  We show that the defenses proposed so far all fit under one of our defense strategies. We also explain how attack graphs can be constructed and point to this as promising future work for tool designers.
\end{abstract}


\blfootnote{Accepted to IEEE International Symposium on High-Performance Computer Architecture (HPCA), 2021}

\section{Introduction} \label{sec:intro}

In computer systems, hardware resources like memory, buses, caches and functional units are often shared among different processes and threads. This sharing increases the utilization of resources. However, preventing a secret from being leaked via shared resources is a fundamental and challenging security problem.

Memory isolation plays a key role in preventing information leakage. An application should not be able to read the memory of the kernel or another application. Memory isolation is usually enforced by the operating system, to allow multiple applications to run simultaneously on the shared hardware resources without information leakage. It is also enforced by the Virtual Machine Monitor to provide isolation between different virtual machines.

Recently, speculative execution attacks, e.g.,  Spectre \cite{kocher2018spectre},
Meltdown \cite{lipp2018meltdown}, Foreshadow \cite{van2018foreshadow}, Foreshadow-NG \cite{weisse2018foreshadow} and Lazy-FP \cite{stecklina2018lazyfp} attacks and their variants are proposed to breach the memory isolation by using a covert channel to exfiltrate a secret obtained
 illegally under speculative execution. 
 For example, Spectre breaches the memory isolation 
 provided within a user application, while Meltdown breaches the memory isolation between the kernel and a user application. Foreshadow breaches the isolation of Intel SGX secure enclaves. Foreshadow-OS and Foreshadow-VMM breach the isolation 
 provided by the Operating System and the Virtual Machine Monitor, respectively. All of these attacks leverage the speculative execution feature of modern processors, transferring the security-critical information to  micro-architecture state 
 observable by an unprivileged attacker
 through a covert channel. 
 Unfortunately, while new attack variants are continuously being discovered, we do not have a systematic way to characterize these attacks and reason about them.  The attack graph model we propose serves this goal.

While both industrial and academic solutions have been proposed to defend against speculative execution attacks \cite{kocher2018spectre, yan2018invisispec, khasawneh2018safespec, saileshwar2019cleanupspec, weisse2019nda, fustos2019spectreguard, barber2019specshield, kirianskydawg, taram2019context, li2019conditional, sakalis2019efficient, schwarz2020context, sabc:ojogbo2020secure, yu2019speculative}, there is currently no systematic way to show if these defenses can defeat speculative attacks, and why.  We show that our attack graph model can explain why a defense will work.

The key questions answered in this paper are:
\ding{172} How can we systematically model the essential common characteristics of speculative execution attacks and reason about them?
\ding{173} What defense strategies can be derived from the new models?
\ding{174} Are the recently proposed defenses effective against these speculative attacks?

Our key contributions in this paper are:
\begin{itemize}
\item We define a new attack graph model to systematically capture the critical operations in speculative execution attacks. 
\item We theoretically prove that a missing edge in an attack graph is equivalent to a race condition, which is one of the root causes of speculative attacks.
\item We define the new concept of ``security dependencies'', which must be observed by the hardware in addition to data dependencies and control dependencies. We show that a missing security dependency is equivalent to a missing edge in an attack graph, capable of causing a security breach.
\item Our model shows that although attacks may look similar, e.g. the Spectre-type and Meltdown-type attacks, they are actually quite different in the sense that Meltdown-type attacks have to be investigated through intra-instruction micro-architectural dependencies, while Spectre-type attacks only need to consider inter-instruction dependencies. This can simplify tool development for finding attack graphs and vulnerabilities that can be exploited in attacks.
\item We derive new defense strategies from our attack graph model. These enable us to systematically explain why a defense will or will not work. We also show that all currently proposed defenses, from both industry and academia, can be modelled by our defense strategies.
\item We show the benefits of our new model for future research for tool creation, discovering new attacks and finding new defenses.
\end{itemize}

\section{Background} \label{sec:bg}

\subsection{Speculative Attacks}\label{sec:bg:spectre}

Speculative execution vulnerabilities affect most modern processors. They exploit speculative execution, Out-of-Order (OOO) execution, hardware prediction and caching -- all essential features for speeding up program execution. They allow an unprivileged adversary to bypass the user-kernel isolation or user-defined boundaries. In a speculative execution attack, a speculation window is induced to allow transient instructions that illegally access a secret, then perform some micro-architectural state changes which essentially ``send out the secret'' so that it can be observed by the attacker. Upon detecting mis-speculation, architectural state changes are discarded, but some micro-architectural state changes are not -- thus leaking the secret.

We give a top-down description of a speculative attack in Section \ref{sec:overview} and a detailed discussion of the Spectre and Meltdown attacks in Section \ref{sec:TSG}. We list the first 13 published attacks and their impacts in Table \ref{tab:bg:attacks}. Later, in Section \ref{sec:application} and Table \ref{tab:TSG:AAnodes}, we also consider the newer attack variants.


\begin{table}[]
\centering
\caption{Speculative attacks and their variants.}\label{tab:bg:attacks}
\resizebox{\linewidth}{!}{
\scriptsize{
\begin{tabular}{|m{2.5cm}|m{2cm}|m{2cm}|}
\hline
\textbf{Attack}                 & \textbf{CVE}   & \textbf{Impact}                                        \\ \hline
Spectre v1 \cite{kocher2018spectre}   & CVE-2017-5753  & Boundary check bypass                                  \\ \hline
Spectre v1.1  \cite{kiriansky2018speculative}       & CVE-2018-3693  & Speculative buffer overflow                            \\ \hline
Spectre v1.2  \cite{kiriansky2018speculative}       & N/A            & Overwrite read-only memory                             \\ \hline
Spectre v2 \cite{kocher2018spectre}  & CVE-2017-5715  & Branch target injection                                \\ \hline
Meltdown (Spectre v3) \cite{lipp2018meltdown}           & CVE-2017-5754  & Kernel content leakage to unprivileged attacker        \\ \hline
Meltdown variant1 (Spectre v3a) \cite{meltdown3a} & CVE-2018-3640  & System register value leakage to unprivileged attacker \\ \hline
Spectre v4 \cite{Spectrev4}          & CVE-2018-3639  & Speculative store bypass, read stale data in memory    \\ \hline
Spectre RSB  \cite{koruyeh2018spectre}   & CVE-2018-15572 & Return mis-predict, execute wrong code                               \\ \hline
Foreshadow (L1 Terminal Fault)\cite{van2018foreshadow}& CVE-2018-3615  & SGX enclave memory leakage                             \\ \hline
Foreshadow-OS \cite{weisse2018foreshadow} & CVE-2018-3620  & OS memory leakage                                      \\ \hline
Foreshadow-VMM \cite{weisse2018foreshadow} & CVE-2018-3646  & VMM memory leakage                                     \\ \hline
Lazy FP \cite{stecklina2018lazyfp} & CVE-2018-3665  & Leak of FPU state                                      \\ \hline
Spoiler \cite{islam2019spoiler}    & CVE-2019-0162  & Virtual-to-physical address mapping leakage            \\ \hline
\end{tabular}
}
}
\end{table}

\subsection{Industry Defenses Implemented}

 Table \ref{fig:bg:defenses-industry} shows some industry defenses that have been implemented to mitigate some speculative attacks.

\begin{table}
  \centering
  \caption{Industrial defenses against speculative attacks.} \label{fig:bg:defenses-industry}
  \includegraphics[width=\linewidth]{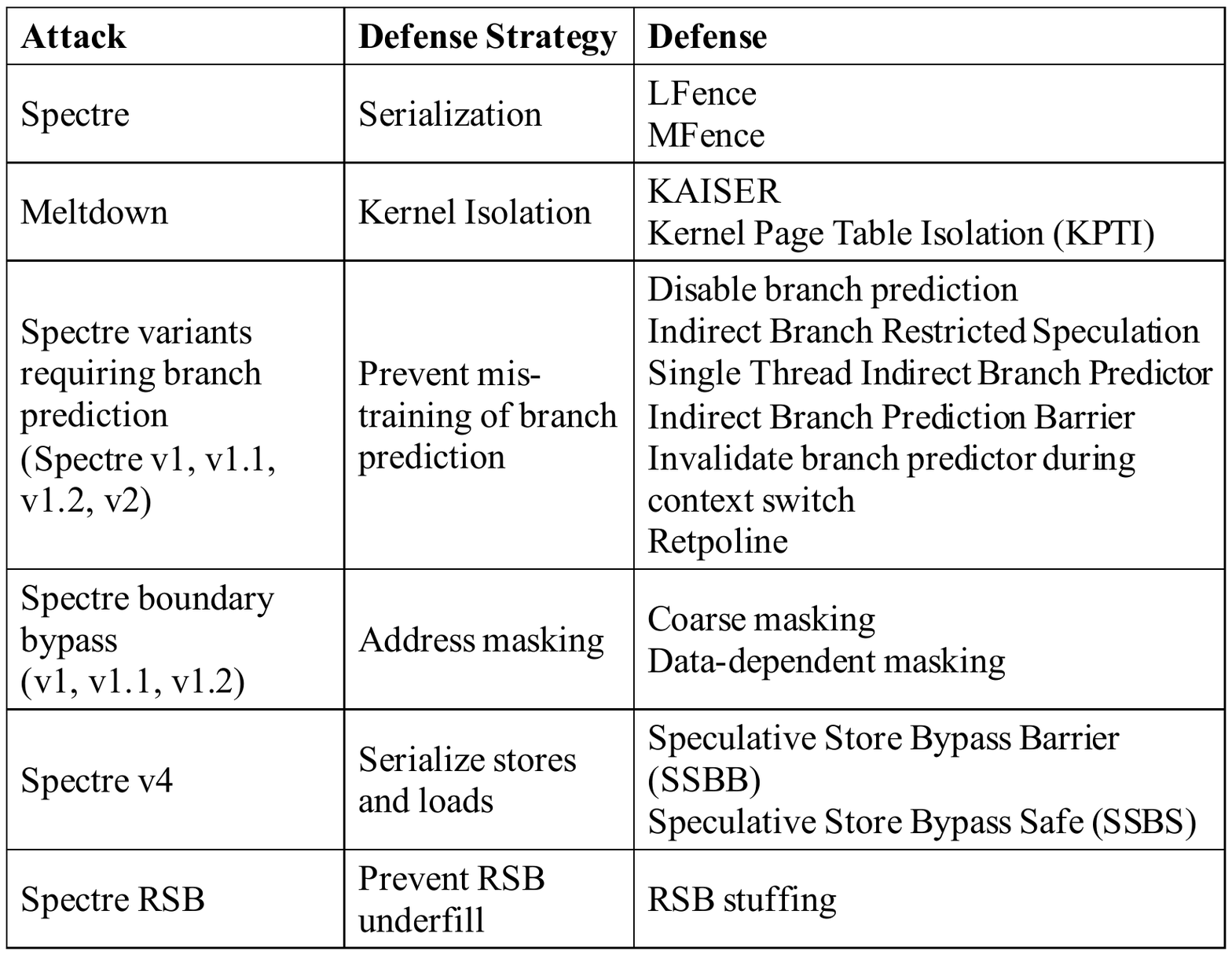}
  \vspace{-15pt}
\end{table}

\bheading{Fences.} Fences, including LFENCE and MFENCE \cite{IntelLFENCE}, are placed before memory operations to serialize the program execution and prevent speculative execution. 

\bheading{Kernel Isolation.} KAISER (Kernel Address Isolation to have Side-channels Efficiently Removed) and its Linux implementation named Kernel Page Table Isolation (KPTI) isolate user-space memory from kernel space to prevent Meltdown attacks, by unmapping kernel pages from user-space \cite{KPTI}.

\bheading{Prevent Mis-training.} As many Spectre variants (v1, v1.1, v1.2, v2) leverage the mis-training of the branch predictors, Intel, AMD and ARM have proposed defenses to prevent mis-training, e.g., Indirect Branch Restricted Speculation (IBRS), Single Thread Indirect Branch Prediction (STIBP) and Indirect Branch Predictor Barrier (IBPB). Some AMD CPUs allow invalidating branch predictor and Branch Target Buffer (BTB) on context switches \cite{AMD-Spectre}.


\bheading{Retpoline.} Retpoline is a method where indirect branches, which use potentially poisoned BTBs, are replaced by return instructions that use the return stack.

\bheading{Address Masking.} To address the problem of software-defined boundary bypass, the V8 JavaScript engine and the Linux kernel implement software address masking by forcing the accessed memory to be within the legal range \cite{kiriansky2018speculative}. 

\bheading{Industrial Defenses against Other Specific Variants.} ARM implemented Speculative Store Bypass Barrier (SSBB) and proposed Speculative Store Bypass Safe (SSBS) to avoid speculative store bypass. Intel implemented Return Stack Buffer (RSB) stuffing, e.g., inserting interrupts to increase call depth, to defend against the Spectre-RSB attack. Intel also announced a silicon-based solution, i.e., the next-generation Cascade Lake processor \cite{intelmitigation}.

\bheading{Academia defenses.}
Recent defenses against speculative attacks have also been proposed in academia, e.g., Context-sensitive fencing \cite{taram2019context}, Secure automatic bounds checking \cite{sabc:ojogbo2020secure}, SpectreGuard \cite{fustos2019spectreguard}, NDA \cite{weisse2019nda}, ConTExT \cite{schwarz2020context}, Specshield \cite{barber2019specshield}, STT \cite{yu2019speculative}, DAWG \cite{kirianskydawg}, InvisiSpec \cite{yan2018invisispec}, Safespec \cite{khasawneh2018safespec}, Conditional Speculation \cite{li2019conditional}, Efficient invisible speculative execution \cite{sakalis2019efficient} and CleanupSpec \cite{saileshwar2019cleanupspec}. We discuss and model them in Section \ref{sec:application}.

\subsection{Cache Timing Channels}
A speculative attack usually includes a covert or side channel attack to leak out the sensitive secret, and a cache covert-channel is typically used.  Hence, we need to understand how cache covert channels work. Cache timing channels can be classified, based on ``hit'' or ``miss'', ``access'' or ``operation''. The access-based attacks leverage the difference in timing between a hit and a miss to indicate whether a specific cacheline is present or absent in the cache, based on a single memory access.
The operation based attacks leverage the time difference for a whole operation, e.g., an encryption operation, which depends on the cache hits or misses encountered during the execution of the operation.

\bheading{Hit and access based channel, \eg, Flush-reload channel \cite{yarom2014flush}.} The initial state of the cacheline is set to absent by a \textit{clflush} instruction. Then, if the insider-sender does not use the shared cacheline, the check by the attacker-receiver after waiting a while, will still find the cacheline absent (indicated by a slow cache miss). If the insider-sender does use the cacheline, then the attacker-receiver will find the cacheline present (indicated by a fast cache hit). 






\bheading{Miss and access based channel, \eg, Prime-probe channel \cite{gullasch2011cache}.} The attacker first loads his own data to fill the cache. After waiting for the insider-sender to do something, the attacker checks if the cachelines are now absent, i.e., a slow cache miss, because the insider-sender has evicted the attacker's cachelines.


There are also hit and operation based channels, \eg, cache collision channel \cite{bonneau2006cache}, and miss and operation based channels, \eg, Evict-time channel \cite{osvik2006cache}.

The Flush-Reload attack is faster and less noisy than the other cache covert channel attacks. They are used as the default covert channels in most speculative attacks. They do require the sender and receiver to have some shared memory.

In the rest of the paper, without loss of generality, we also assume that the Flush-Reload cache covert channel is used in the speculative attacks. Our models can also apply to the prime-probe channel, and other non-cache-based covert channels, with minor changes.

\section{Overview of Speculative Attacks} \label{sec:overview}

In speculative attacks, the micro-architectural feature attacked is speculative execution, in concert with out-of-order (OoO) execution. Out-of-order execution allows instructions to be executed once their data operands are ready, i.e., when their data dependencies are resolved. This does not need to be in sequential program order.  When an instruction is issued, it is placed into a Re-Order Buffer (ROB), with the oldest instruction at the head of the ROB and the youngest at the tail. Once an instruction's data dependencies are resolved, the instruction can be executed. It is not committed (i.e., retired) until it reaches the head of the ROB, i.e., instructions are retired in program order.  This out-of-order execution speeds up instruction processing.

Speculative execution is a performance optimization feature that occurs when the hardware micro-architecture predicts that a certain execution path will be taken on a conditional branch, or that an instruction will not cause an exception. If the prediction is correct (which is most of the time), performance is improved.  However, if the prediction is wrong, then the hardware rolls back the architecturally-visible state to where it was before the speculation window started. The speculatively executed instructions appear as if they were never executed, i.e., the mis-predicted instructions are aborted or squashed. While processors implement speculation correctly as defined, the micro-architectural state is not always rolled back completely, as this is not supposed to be an architecturally-visible state. In particular, caches are considered micro-architecture, and are not rolled back.

Although the exact workflow of a speculative execution attack may vary, on a high-level, they consist of two parts:
\textbf{\textit{
\begin{enumerate}[label=(\Alph*),nolistsep]
  \item Secret Access: A micro-architectural feature transiently enables the illegal access of a piece of sensitive data.
  \item Secret Send or Covert Channel: The sensitive data is transformed into  micro-architectural state that can be observed by an attacker.
\end{enumerate}
}}

\noindent \textbf{Definition 1: an illegal access} is a data or code access that is performed before the required authorization is completed that indicates that the access is allowed. The required \textbf{authorization} is the operation checking if the performer is allowed to access the data, or execute the code. Authorization can be in different forms, e.g., a hardware privilege level check, a software array bounds check or a store-load address dependency check.

Since our definition of ``authorization'' is broader than the standard user-supervisor-hypervisor access checking, we give examples to illustrate. In the Meltdown attack, the attacker tries to read a memory line before the hardware page-privilege check that indicates the performer of the memory access has kernel privilege. In the Spectre v1 attack, the illegal access is reading out-of-bounds memory at the user level. The array bounds check (authorization) is the condition checking in a conditional branch instruction. Although the memory access is within the legal program address space, we call it an illegal access because the software-enforced array bounds checking has not been completed. In the Spectre v4 attack (store-load dependency), we call the load operation an illegal access if it reads stale data before the authorization completes that says the load address is not the same as the address of a previous store operation that is still sitting in the store buffer and its contents have not been written back to the cache.

To defend against speculative attacks, one must prevent either part A (Secret Access) or part B (Secret Send or Covert Channel). By preventing A, the access to secrets, there are no secrets to leak through any covert or side channel. By preventing B, any secrets present cannot be exfiltrated, nor can secrets obtained through means other than speculation, be leaked. However, there can be many types of covert channels, and stopping all of them is not possible. Although computer architecture papers have focused on preventing cache covert channels, we believe this is only a near-term solution, since the attacker can easily find other covert channels to leak the secret information. We do not want to exclude these other covert channels. Hence, in this paper, we focus on modeling the illegal access to secrets through speculative execution, and having our attack model capable of modeling any covert channel.

Parts A and B can be broken down into the following finer-grained attack steps that are critical to the success of a speculation attack. There are 5 steps for an actual attack, and 6 if we count step (0) where the attacker finds the location of the desired secret. This is usually done earlier, before the actual speculative attack.

\begin{itemize}
\item \textbf{(Step 0):} Know where the secret is.
\item \textbf{Step 1 (Setup):} Receiver (a) establishes a micro-architectural covert-channel, e.g., by flushing out cachelines, and (b) sets up for illegal access, e.g., by mis-training the branch predictor.
\item \textbf{Step 2 (Authorization):} The instruction performing the authorization for the subsequent memory or register access is \emph{delayed}, thus triggering the start of a speculative execution window.  If the authorization turns out to be negative, then the instructions executed speculatively are called transient instructions since they are squashed.  If authorization turns out to be positive, then the instructions executed speculatively are committed.
\item \textbf{Step 3 (Secret Access):} Sender (illegally) accesses the secret.
\item \textbf{Step 4 (Use Secret and Send Secret):} Sender transforms the secret into a micro-architectural state that survives mis-speculation.
\item \textbf{Step 5 (Receive Secret):} Receiver retrieves micro-architecture state (the transformed secret) through the covert-channel.
\end{itemize}

\noindent Steps 0, 1(b), 2 and 3 form part A. Steps 1(a), 4 and 5 form part B.

\section{Attack Graph and Security Dependency}\label{sec:TSG}


We now look at specific speculative execution attacks, and model the flow of relevant operations that occur, to help reason about the attacks, and identify the root causes of their success. In Section \ref{sec:TSG:motivation}, we model the Spectre v1 attack as a flow graph, and confirm that it follows the five steps we identified in Section \ref{sec:overview}. This motivates us to define an attack graph in Section \ref{sec:TSG:graph}, as a topological sort graph (TSG), which enables us to formally prove necessary and sufficient conditions for a race condition to occur, which we identify as a root cause of the success of speculative attacks. In Section \ref{sec:TSG:dependency}, we propose the concept of \bheading{security dependency}, and equate this with inserting  a missing edge between two operations in the attack graph that will defeat the attack. In Section \ref{sec:TSG:Meltdown}, we model the Meltdown attack with an attack graph, and in Section \ref{sec:TSG:Other}, we show that our attack graph models can be extended to all attack variants. 

\subsection{Example: Spectre v1 Attack}\label{sec:TSG:motivation}



Spectre attacks exploit the \textit{transient instructions} which can execute during a speculative execution window. On a mis-speculation, the transient instructions are aborted and all architectural-level side effects are rolled back. However, not all micro-architectural state changes are rolled back.

Listing \ref{list:Spectre} shows an example of the Spectre v1 attack, bypassing the array bounds checking, thus reading arbitrary content that is not allowed, then sending the transformed secret out using a Flush-Reload cache side-channel.

\begin{lstlisting}[caption={Code snippet of the Spectre v1 attack to bypass array bounds checking, using the Flush-Reload channel.}, captionpos=b, label={list:Spectre}, basicstyle=\footnotesize, numbers=left,xleftmargin=2em,framexleftmargin=1.5em]
// Establish channel by flushing shared Array_A accessible to attacker
int T[256]
char Array_A[256*4K]
clflush(Array_A)

// Train the branch predictor to predict not taken
train_branch_predictor()

mov rbx, Array_A
mov rcx, Array_Victim
mov rdx, Secret_Location in Array_Victim

// if (x < Array_Victim_Size)
// y = Array_A[Array_Victim[x] * 256];

// rdx stores the queried index x and if x >
// Victim_Array_Size, the branch should be taken
cmp rdx, Array_Victim_Size              //Authorization
ja .BRANCH_TAKEN
// Speculative Execution window starts

// Illegal memory access of Secret_Location
mov al, byte [Array_Victim + rdx]       // Access

shl rax, 0xc                            // Use
mov rbx, qword [rbx + rax]              // Send

.BRANCH_TAKEN: ...

// Reload Array_A to recover the secret byte
for(i=0; i<256; i++){
    T[i] = measure_access_time(Array_A + i * 4KB)
}
recovered_byte = argmin_i(T[i]).        //Receive
\end{lstlisting}

Suppose the target victim's secret is located at $Secret\_Location$.
Lines 1-4 prepare the Flush-Reload side channel by flushing the cachelines of Array\_A, which is accessible to the attacker and the victim. In line 7, the attacker trains the branch predictor to always predict not taken. Lines 9 and 10 put the base address of shared Array\_A and private Array\_Victim into registers. Line 11 sets rdx such that $Array\_Victim$[rdx] points to $Secret\_Location$. Note that $Array\_Victim$ itself may not have sensitive data, but rdx exceeds the length of $Array\_Victim\_Size$ and refers to the secret.

Lines 13-14 show the high-level C code of the assembly code in lines 16-26. This is the crux of the Spectre v1 attack. Line 18 is an array bounds checking, where rdx is compared to $Array\_Victim\_Size$. However, if getting $Array\_Victim\_Size$ is delayed, e.g., not in the cache, the branch predictor will predict the branch in line 19 as not taken because the attacker has mistrained the predictor in line 7. Line 23 illegally reads the secret into the low-order byte of register rax. Line 25 transforms the secret into an index of Array\_A (where each value of secret refers to a new page, to avoid unwanted prefetching). Line 26 exfiltrates the secret by accessing an entry at Array\_A indexed by the secret, thus changing the state of the cacheline from absent to present for the Flush-Reload attack. When $Array\_Victim\_Size$ finally arrives and the comparison is done, the processor realizes the mis-prediction in line 19, and discards the values in rax and rbx. However, the cache state is not rolled back, and the attacker can obtain the secret by checking which entry of $Array\_A$ has been fetched (lines 30-34), since this entry gives a cache hit.

We model the Spectre v1 attack in Figure \ref{fig:TSG:spectre}. This is the first example of an Attack Graph. Here, the nodes are instructions and the links are data or control dependencies.  The dotted arrows represent the speculative execution path.



Figure \ref{fig:TSG:spectre} follows the program flow in Listing \ref{list:Spectre}. It also follows the five steps outlined in Section \ref{sec:overview}. First, the receiver sets up the covert channel by flushing Array\_A and mis-training the branch predictor, such that the branch prediction will predict ``not taken'' (step 1). During the program execution, the branch stalls as the branch condition has not been resolved (the authorization operation, step 2). The branch predictor allows the speculative load of the secret (``Load S'') to be performed (step 3), bypassing the program-defined authorization. After the secret is obtained, the sender exfiltrates it by fetching a secret-related entry in Array\_A (step 4). Finally, the receiver retrieves the secret by reloading entries in Array\_A and measuring the access time (step 5). A short access time indicates that the entry in Array\_A indexed by secret has been fetched into the cache. Some key observations and insights are:

\bheading{Speculative execution window.} Once the branch stalls as the condition has not been resolved, the (possibly incorrect) instructions are speculatively executed in a speculative window. The speculative window is marked by the red 
dashed block in Figure \ref{fig:TSG:spectre}. The speculative window starts from the issue of the first speculative (or transient) instruction until the branch condition ultimately resolves. If mis-predicted, the speculated instructions are squashed; otherwise, they are committed, improving the performance.

\bheading{Speculated Operations Race with the Authorization.} The speculatively executed instructions and the branch resolution (i.e., the authorization) are performed concurrently. In particular, whether the two memory load operations or the branch resolution finishes first, is non-deterministic. Hence, there are two race conditions between ``Load S'' (secret access), ``Load R'' (micro-architecture state change) and ``Branch resolution'' (software authorization). 

\bheading{The race condition allows unauthorized access.} The memory operation ``Load S'' in the speculative window race can be outside the software-defined boundary. Thus it is an unauthorized or illegal memory access.

\bheading{The race condition is due to a missing security dependency.} The race condition is because of a missing security dependency (formally defined in Section \ref{sec:TSG:dependency}) between branch resolution and ``Load S''. It is neither a data dependency nor a control dependency, but a new dependency to decide when an operation can be executed. This missing security dependency was first pointed out by Lee \cite{Micro51Keynote} as the root cause of speculative execution attacks, since the ``No Access without Authorization'' security principle is violated.

\subsection{Attack Graph and Races} \label{sec:TSG:graph}

We define an Attack Graph to extend and formalize the connection between a race condition and a missing dependency. We define an attack graph as a Topological Sort Graph (TSG), a directed acyclic graph with edges between vertices representing orderings.



\begin{figure}[t]
\centering
\includegraphics[width=0.4\columnwidth]{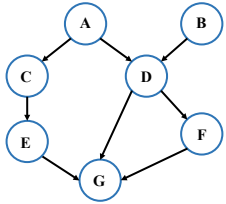}
\caption{An example of Topological Sort Graph (TSG).} \label{fig:TSG:example}
\vspace{-15pt}
\end{figure}

\bheading{A vertex} in a TSG represents an operation, e.g., accessing a memory line, flushing a cacheline or comparing a memory address to a bound. Figure \ref{fig:TSG:example} shows an example of a TSG.

\bheading{A directed edge} in the TSG represents a dependency of two vertices. If there is an edge from $u$ to $v$, $u$ happens before $v$.

\bheading{A path} is a sequence of edges that connects vertices. All paths considered in this paper are directed paths.

\bheading{An ordering} of vertices in a TSG is an ordered list that contains all vertices in the graph $S=(v_1,v_2...v_n).$ An ordering of vertices in a TSG is \textbf{valid}, if and only if for every directed edge ($v_i,v_j$) from vertex $v_i$ to vertex $v_j$, $v_i$ comes before $v_j$ in the ordering. For example, in Figure \ref{fig:TSG:example}, $S=[A,B,C,D,E,F,G]$ and $S'=[A,C,E,B,D,F,G]$ are both valid orderings. $S''=[A,B,D,E,C,F,G]$ is not a valid ordering.

\bheading{A race condition} exists between vertex $u$ and $v$ in a TSG if there exists two different valid orderings $S_1$ and $S_2$ such that $u$ is before $v$ in $S_1$, and $v$ is before $u$ in $S_2$. Take Figure \ref{fig:TSG:example} as an example, there is a race condition between $D$ and $E$, because $S=[A,B,C,D,E,F,G]$ and $S'=[A,C,E,B,D,F,G]$, but $D$ is before $E$ is $S$ and $D$ is after $E$ in $S'$.

\begin{figure}[t!]
    \centering
    \includegraphics[width=\linewidth]{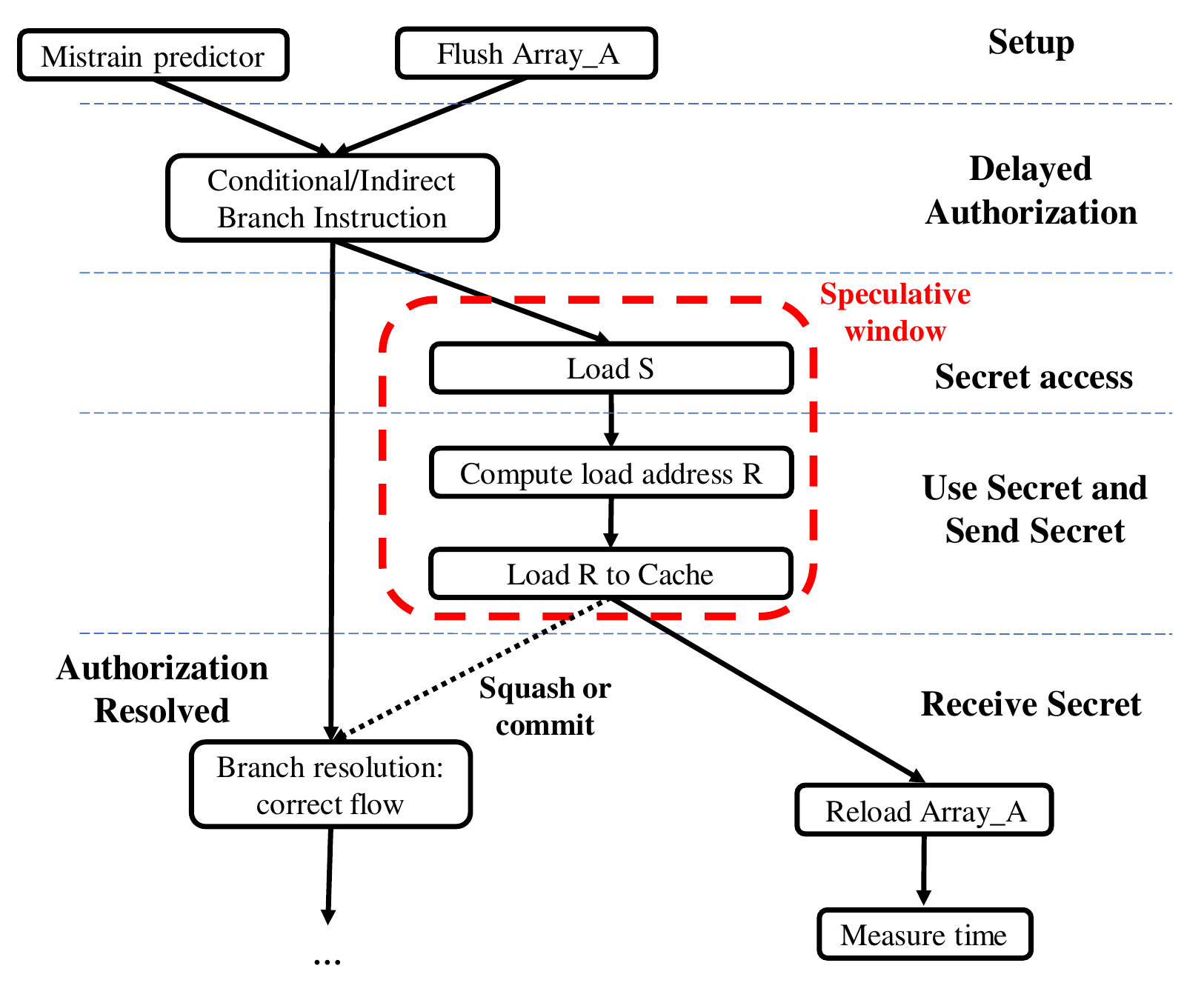}
    \caption{Spectre v1/v2 attacks. The speculative execution window is marked by the red dashed block.
    ``Branch resolution'' marks the completion of the delayed authorization, initiated by the conditional or indirect branch instruction. ``Load S'' (secret-accessing) and ``Load R'' (secret-sending) are unauthorized memory accesses if they bypass ``Branch resolution'' (software-defined authorization).} \label{fig:TSG:spectre}
    \vspace{-15pt}
\end{figure}




We prove the following theorem connecting a race condition with a missing dependency. 

\textbf{\textit{Theorem 1.} For any pair of vertices $u$ and $v$, the two vertices $u$ and $v$ do not have a race condition, if and only if there exists a directed path that connects $u$ and $v$.}

We provide a formal proof in Appendix \ref{sec:appendix}.

Given a directed graph of operations, there are methods to efficiently check whether there is a path between two vertices \cite{aggarwal1989parallel}, using depth-first search. If none exists, there is a race condition between these two operations.

To build an attack graph, all branch, memory access (load and store) and arithmetic instructions need to be included in the graph. 
Data dependencies are shown as existing edges in the attack graph. Since not all operations and race conditions in a computation are relevant, we define four types of vertices that must be represented in an attack graph: 

\bheading{Authorization Operations.} The victim or covert sender's authorization operations are nodes in the attack graph, representing the permission checking and other forms of authorization, e.g., array bounds checking by a conditional branch in the user program.

\bheading{Sender's Secret Access Operation.} The sender's secret access operation is a node in the attack graph, representing access to the secret. For example, this is the out-of-bounds memory access (Load S) in Figure \ref{fig:TSG:spectre}.

\bheading{Sender's Send (Micro-architecture State Change) Operation.} A node where the sender manipulates the micro-architecture state according to the secret, e.g., the memory access ``Load $R$ to cache'' for the Flush-Reload cache covert channel.

\bheading{Receiver's Secret Access Operation.} This is a node representing the retrieval of the secret from the micro-architecture covert-channel. For example, it is a memory read and access time measuring operation in a cache Flush-Reload or Prime-Probe covert channel.

\subsection{Security Dependency} \label{sec:TSG:dependency}

\textbf{Definition 2: A security dependency} of operation $v$ on operation $u$ is an ordering of the two operations such that $u$ must be completed before $v$, in order to avoid security breaches. Operation $u$ is typically a security protection operation, which we call an authorization operation
in this paper. Operation $v$ is typically an illegal access of data or code.

Following the ``No access without Authorization'' \cite{Micro51Keynote} security principle strictly means that the authorization has to be completed \textit{before} the protected data access or code execution. This introduces a security dependency between authorization and data access (or code execution), which prevents the race condition that is the root cause of speculative attacks. Like the well-studied data dependencies and control-flow dependencies, which must be followed to ensure correct program execution, security dependencies must be followed to enforce the security of program execution.

However, as we will show in Section \ref{sec:app:defense}, some security-performance tradeoffs can be made that still prevent attacks from succeeding, by making sure that even if the secret is fetched, it is prevented from being used or exfiltrated out to an attacker-receiver.

\subsection{Modeling Meltdown Attacks} \label{sec:TSG:Meltdown}

We show a code snippet of the Meltdown attack in Listing \ref{list:Meltdown}. The front and back parts of the Meltdown attack are similar to the Spectre v1 attack in setting up the covert channel (step 1, lines 1-4), using and sending out the secret (step 4, lines 12-14) and testing the covert channel (step 5, lines 16-20). The main difference is in line 10, which accesses supervisor memory and should cause an exception. If the exception is delayed, a speculative window is triggered. There is a race condition between the speculative execution of lines 10, 13-14 with the raising of the exception in line 10.


\begin{lstlisting}[caption={A code snippet of the Meltdown attack.}, captionpos=b, label={list:Meltdown}, basicstyle=\footnotesize, numbers=left,xleftmargin=2em,framexleftmargin=1.5em]
// Establish the covert channel by flushing Array_A
int T[256]
char Array_A[256*4K]
clflush(Array_A)				    \\Setup

mov rbx, Array_A
mov rcx, Security_Critical_Memory_Addr

// Illegal memory access
mov al, byte [rcx]				    \\Authorize and Access

// Speculatively execute the transient instructions
shl rax, 0xc					    \\ Use
mov rbx, qword [rbx + rax]			\\Send

// Reload Array_A to recover the security-critical byte
for(i=0; i<256; i++){
    T[i] = measure_access_time(Array_A + i * 4KB)
}
recovered_byte = argmin_i(T[i])		\\Receive
\end{lstlisting}

\begin{figure}[h]
    \centering
    \includegraphics[width=\linewidth]{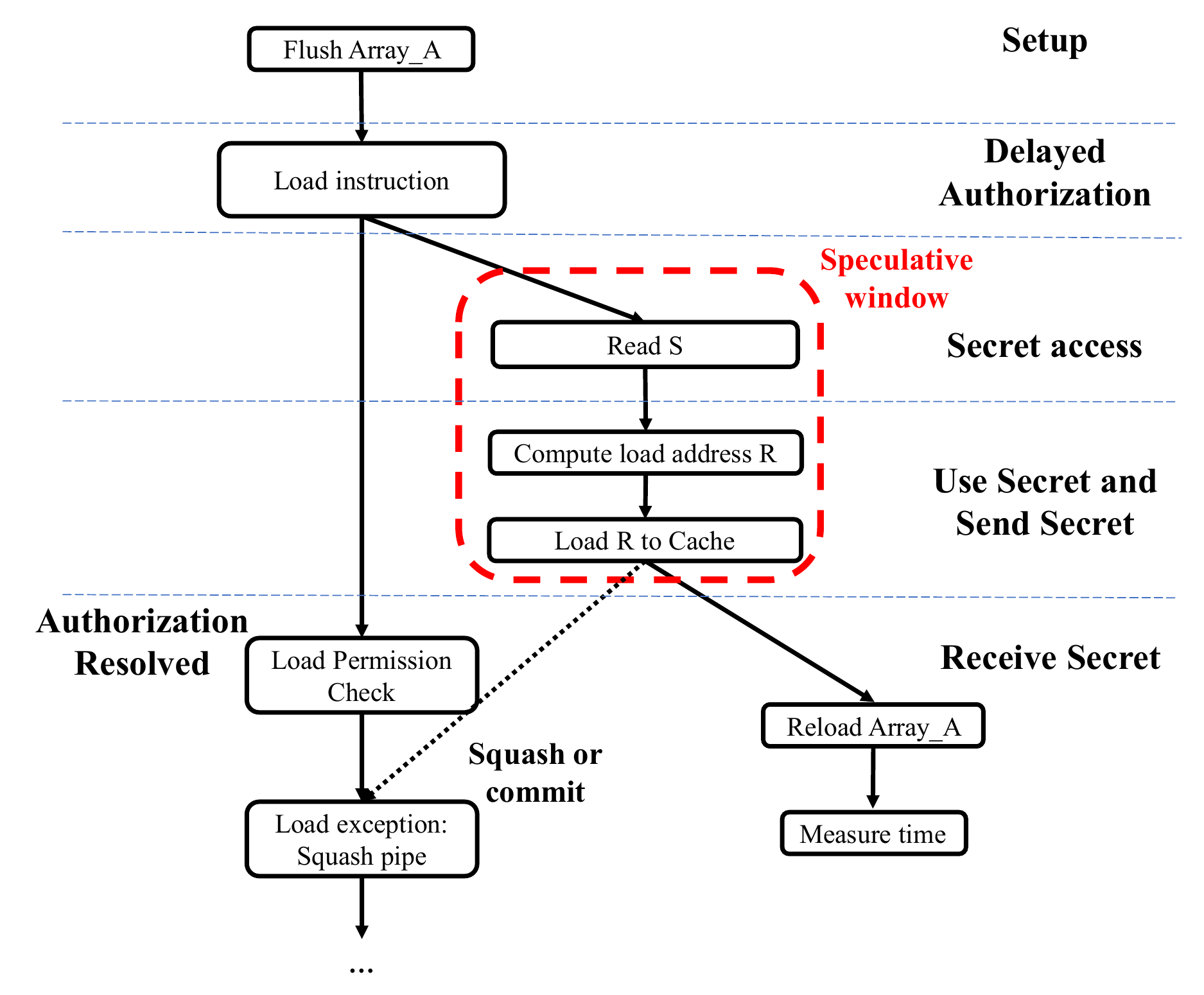}
    \caption{TSG model of the Meltdown attack.} \label{fig:TSG:meltdown-microarch}
    \vspace{-10pt}
\end{figure}

\emph{Our insight is that in the Meltdown type of attacks, the Authorization and the secret Access are actually the same instruction - a memory load instruction.  Hence, we need to look within this instruction and model its micro-architectural operations that may race with each other.}

The attack graph of Meltdown in Figure \ref{fig:TSG:meltdown-microarch} is similar to that for Spectre in Figure \ref{fig:TSG:spectre}, except that this time we show the micro-architecture operations of the load instruction in separate nodes, rather than just a single node for a conditional branch instruction and a separate node for a ``Load S'' instruction. The delayed privilege check  (authorization) triggers the start of speculative execution, allowing the illegal access of the secret in the ``Read S'' operation. It also allows the micro-architectural change of the cacheline from absent to present in the ``Load R to cache'' instruction, which results in a hit on this cacheline, leaking the secret in the Flush-Reload cache covert channel.

\begin{figure*}[t!]
    \centering
    \includegraphics[width=0.8\linewidth]{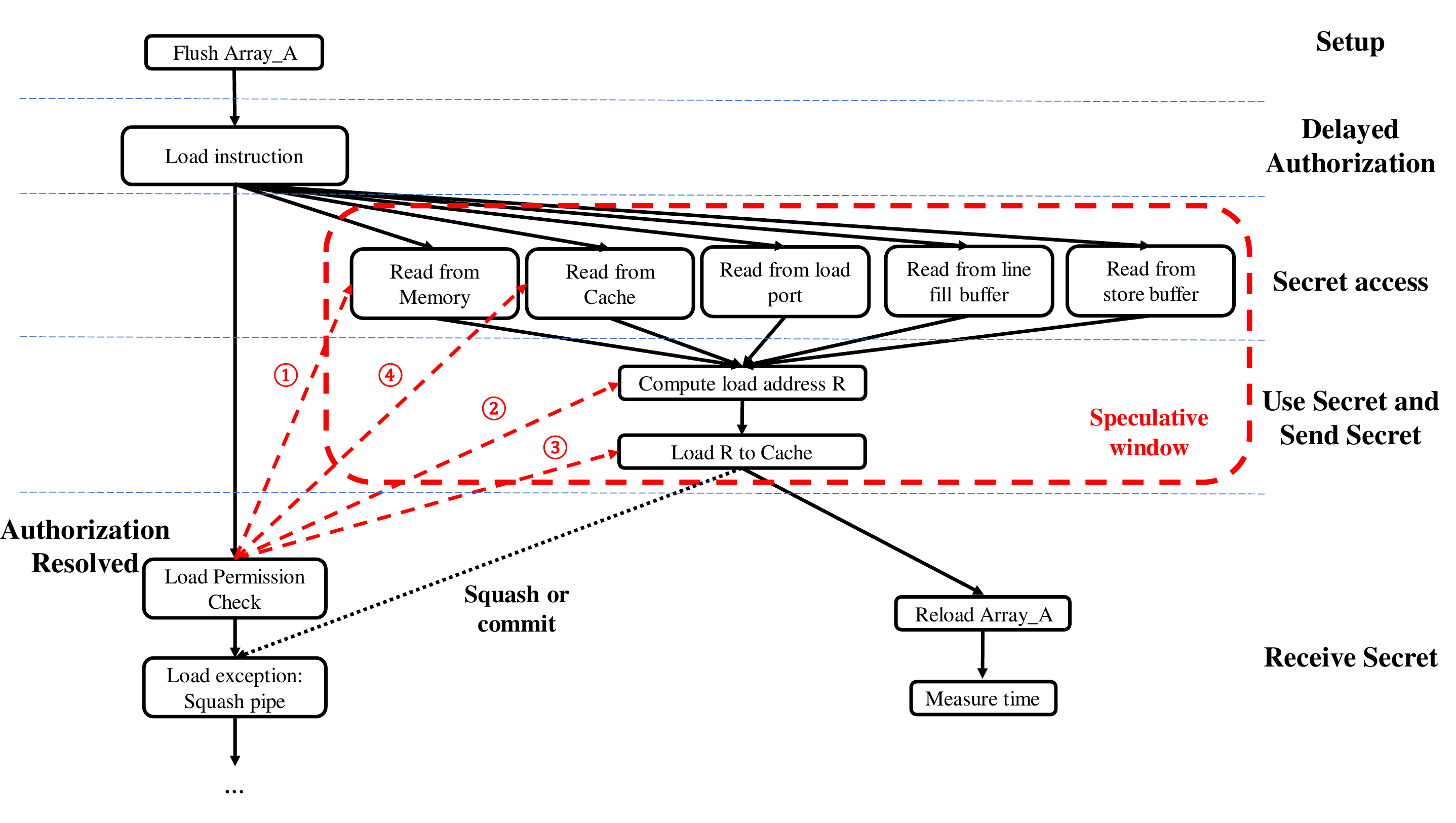}
    \caption{Attack graph model for the Meltdown, Foreshadow and MDS attacks. The source of the secret can be from: the memory (Meltdown), cache (Foreshadow), load port (RIDL), line fill buffer (RIDL and ZombieLoad) or store buffer (Fallout). The red dotted lines indicate different defense strategies that can prevent the attacks from succeeding (discussed in Section \ref{sec:application}).} \label{fig:TSG:meltdown} 
    \vspace{-10pt}
\end{figure*}

\subsection{Modeling Other Attacks} \label{sec:TSG:Other}

Our attack graphs can be generalized to all the speculative attacks, and potentially other micro-architectural security attacks. In Table \ref{tab:TSG:AAnodes}, we summarize the authorization nodes and illegal access nodes for all the speculative attack variants, to illustrate that our attack graph model can be generalized. We describe these attack variants below, including the newer attacks added at the bottom of Table \ref{tab:TSG:AAnodes}.

\begin{table}[t]
\centering
\caption{Authorization and Access Nodes of Speculative Attacks.}\label{tab:TSG:AAnodes}
\resizebox{\columnwidth}{!}{
\scriptsize{
\begin{tabular}{|m{2cm}|m{2cm}|m{2cm}|}
\hline
\textbf{Attack}   & \textbf{Authorization}   & \textbf{Illegal Access}   \\ \hline
Spectre v1 \cite{kocher2018spectre}   & Boundary-check branch resolution  &  Read out-of-bounds memory \\ \hline
Spectre v1.1  \cite{kiriansky2018speculative} & Boundary-check branch resolution & Write out-of-bounds memory \\ \hline
Spectre v1.2  \cite{kiriansky2018speculative} & Page read-only bit check & Write read-only memory \\ \hline
Spectre v2 \cite{kocher2018spectre}  & Indirect branch target resolution & Execute code not intended to be executed \\ \hline
Meltdown (Spectre v3) \cite{lipp2018meltdown}  &  Kernel privilege check  & Read from kernel memory \\ \hline
Meltdown variant1 (Spectre v3a) \cite{meltdown3a} & RDMSR instruction privilege check & Read system register \\ \hline
Spectre v4 \cite{Spectrev4}          & Store-load address dependency resolution & Read stale data    \\ \hline
Spectre RSB  \cite{koruyeh2018spectre}  & Return target resolution & Execute code not intended to be executed \\ \hline
Foreshadow (L1 Terminal Fault)\cite{van2018foreshadow}& Page permission check & Read enclave data in L1 cache from outside enclave \\ \hline
Foreshadow-OS \cite{weisse2018foreshadow} & Page permission check  & Read kernel data in cache  \\ \hline
Foreshadow-VMM \cite{weisse2018foreshadow} & Page permission check & Read VMM data in cache \\ \hline
Lazy FP \cite{stecklina2018lazyfp} & FPU owner check & Read stale FPU state  \\ \hline
RIDL \cite{van2019ridl} & Load fault check & Forward data from fill buffer and load port \\ \hline
ZombieLoad \cite{schwarz2019zombieload} & Load fault check & Forward data from fill buffer \\ \hline
Fallout \cite{canella2019fallout} & Load fault check & Forward data from store buffer \\ \hline
LVI \cite{van2020lvi} & Load fault check & Forward data from micro-architectural buffers (L1D cache, load port, store buffer and line fill buffer) \\ \hline
TAA \cite{canella2019fallout} & TSX Asynchronous Abort Completion & Load data from L1D cache, store or load buffers \\ \hline
Cacheout \cite{van2020cacheout} & TSX Asynchronous Abort Completion & Forward data from fill buffer \\ \hline
\end{tabular}
}
}
\end{table}

\bheading{\ding{172} The Foreshadow or ``L1 terminal fault'' attacks.}

The Foreshadow type of attacks exploit a hardware vulnerability that allows the attacker, such as in Foreshadow \cite{van2018foreshadow} or Foreshadow-NG \cite{weisse2018foreshadow}, to read a secret from the L1 data cache, instead of from the memory, as in the Meltdown attack. The speculative execution of an instruction accessing a virtual address with page table entry marked not present or the reserved bits set, will read from L1 data cache as if the page referenced by the address bits in the PTE is still present and accessible. The L1 terminal fault attack can be leveraged to breach the SGX isolation, as the speculative load bypasses the extended page table (EPT) protection and the secure enclave protection of SGX and reads secret data from the L1 cache.

Hence, these attacks can be modeled by the same attack graph as for the Meltdown attack, but the attack flow goes down the ``Read from cache'' branch in Figure \ref{fig:TSG:meltdown} instead of the ``Read from memory'' node. The permission check is performed for the present bit or the reserved bit in the page table, which can cause the address translation to abort prematurely.

\bheading{\ding{173} MDS attacks (RIDL, ZombieLoad and Fallout).}

The newer Micro-architectural Data Sampling (MDS) attacks, e.g., Rogue In-Flight Data Load (RIDL) \cite{van2019ridl}, ZombieLoad \cite{schwarz2019zombieload} and Fallout attacks \cite{canella2019fallout}, leverage the hardware mechanisms that allow a load that results in a fault to speculatively and aggressively forward stale data from micro-architectural buffers. These attacks use different micro-architectural buffers as the source for accessing the secret, shown as different attack paths in Figure \ref{fig:TSG:meltdown}: RIDL reads a secret from a load port or line fill buffer, ZombieLoad reads a secret from a line fill buffer and Fallout reads a secret from a store buffer. To model these attacks in the attack graph, we also generalize the ``permission check'' to include the check for hardware faults that may trigger this illegal secret access.

\bheading{\ding{174} Special Register attacks (Spectre v3a and Lazy FP).}

\begin{figure}[t]
    \centering
    \includegraphics[width=\linewidth]{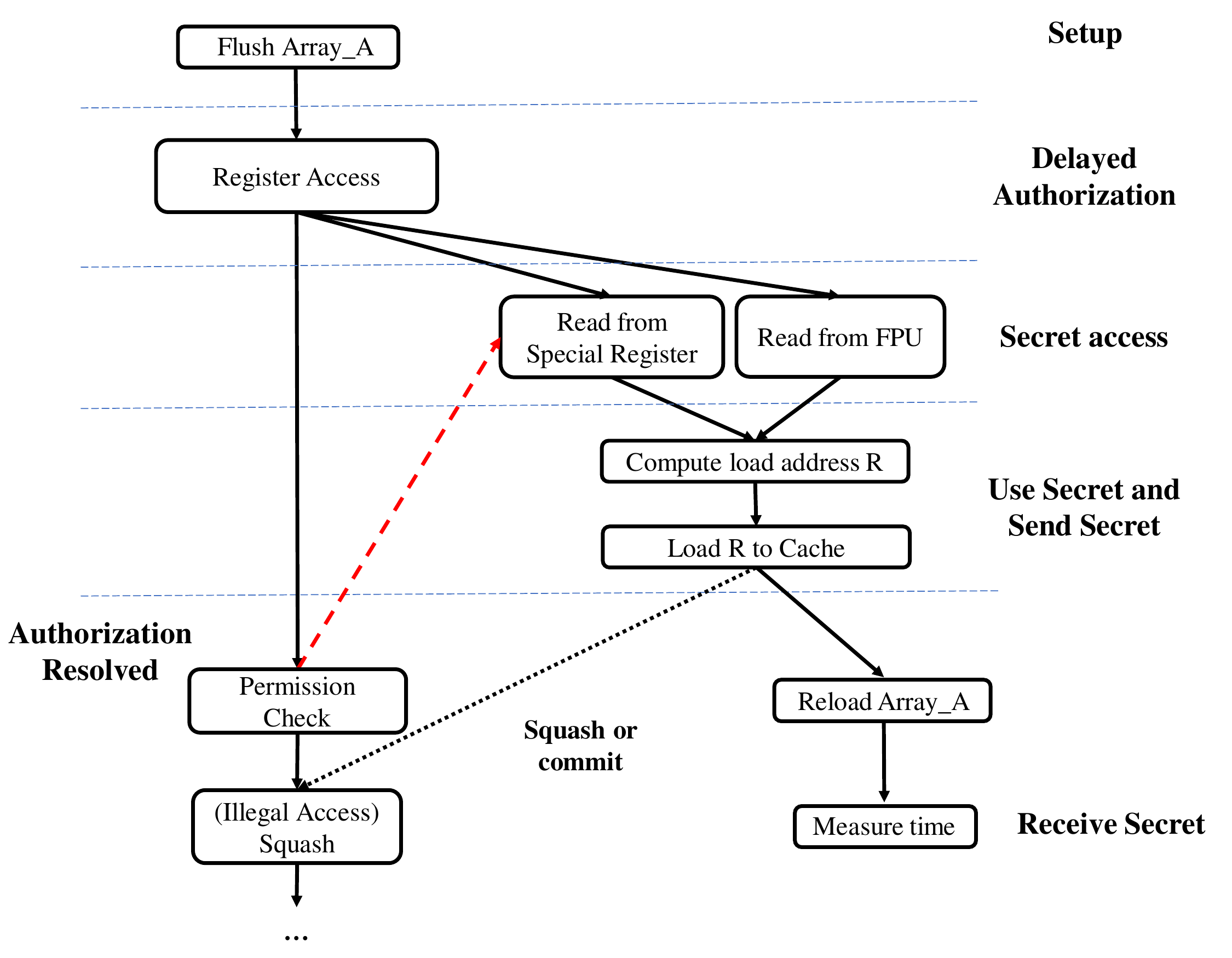}
    \caption{TSG model of special register triggered attacks.} \label{fig:TSG:more-attack-register}
    \vspace{-10pt}
\end{figure}

Another source of secrets is from the reading of special registers, i.e., not the general-purpose registers, rather than reading from the cache-memory system. We model these attacks in Figure \ref{fig:TSG:more-attack-register}, where the illegal access is reading from these registers.

The Spectre v3a (Rogue System Register Read) attack can have a delayed authorization due to privilege checking (for supervisor privilege) taking longer than reading the system register. The implied hardware prediction is that the privilege checking passes, so the system register is accessed speculatively.

In the Lazy FP attack, the floating-point registers are not immediately switched on a context switch, but only switched when a floating-point instruction is actually encountered. Hence, there is a delay in the first floating-point instruction encountered in a new context that can result in speculatively accessing the old values of the floating-point registers of the previous context. We show the missing security dependency as a red arrow from authorization to read register.

\bheading{\ding{175} Indirect branch attack (Spectre v2).}

The Spectre v2 attack mis-trains hardware predictors, e.g., the branch target buffer (BTB), such that the victim speculatively jumps to a wrong address and executes malicious gadgets (i.e., code) that can access and leak a secret. This attack can also be modeled by Figure \ref{fig:TSG:spectre}. The difference with Spectre v1 is that the speculative execution starts because the computation of the target address is delayed, and so the prediction for the target address (BTB) of the indirect branch instruction is used instead. The ``authorization'' of the control flow defined by the indirect branch instruction is completed when the real branch target address is computed and compared with the predicted target address.

\bheading{\ding{176} Memory disambiguation triggered attack (Spectre v4).}

\begin{figure}[t]
    \centering
    \includegraphics[width=\linewidth]{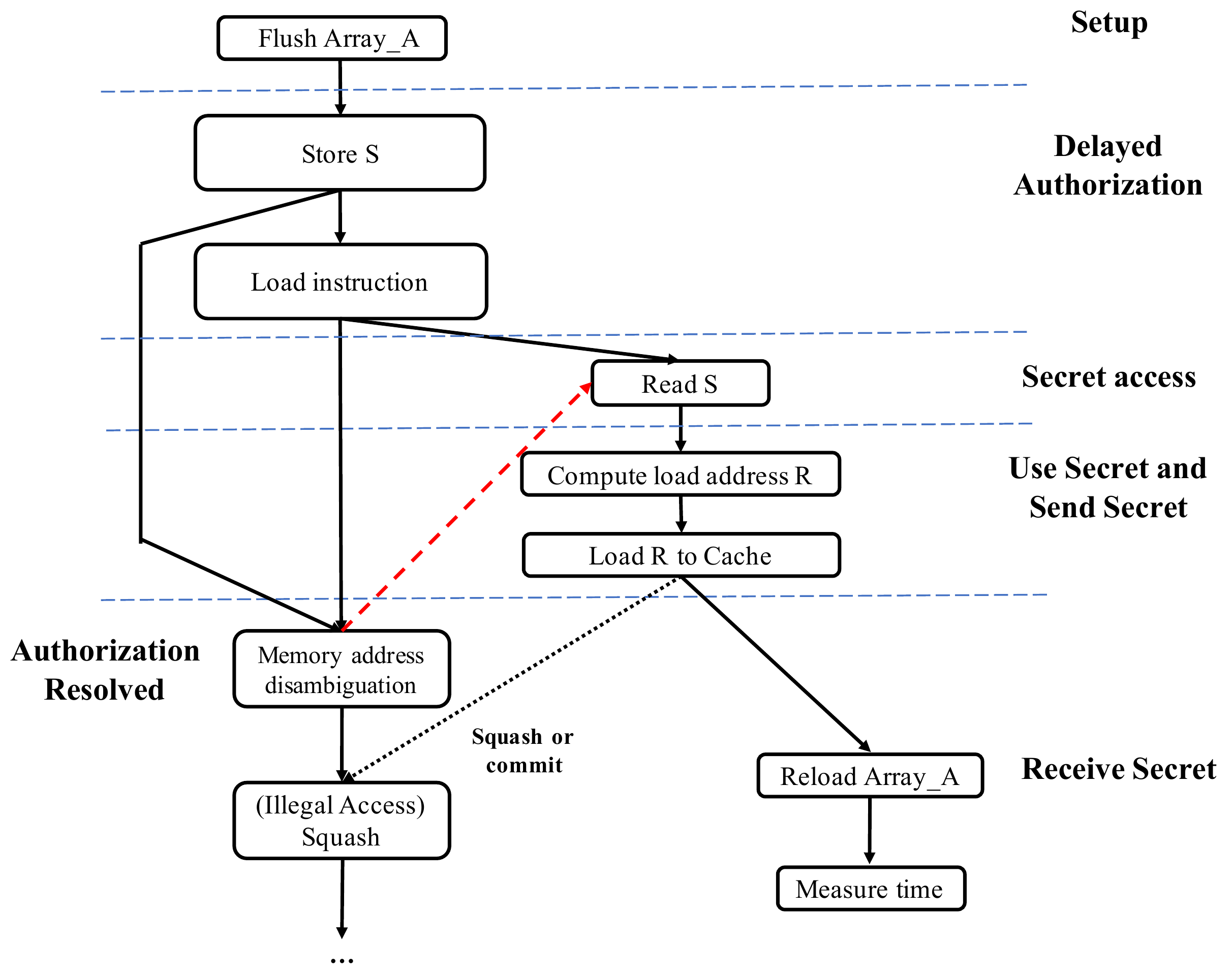}
    \caption{TSG model of memory disambiguation triggered attack.} \label{fig:TSG:more-attack-mem}
    \vspace{-10pt}
\end{figure}

The Spectre v4 (Spectre-STL) attack speculatively reads stale data (secret) that should be overwritten by a previous store. During the speculative load, address disambiguation mispredicts that the load does not depend on a previous store, i.e., the load address is not the same as any of the addresses of store instructions still sitting in the store buffer. We model this attack in Figure \ref{fig:TSG:more-attack-mem}. The authorization is address disambiguation and the illegal access is ``Read S''. A missing dependency is shown as the red dashed arrow from address disambiguation to the illegal access ``Read S''.

\bheading{\ding{177} Load Value Injection (LVI) attack.}

\begin{figure}[t]
    \centering
    \includegraphics[width=\linewidth]{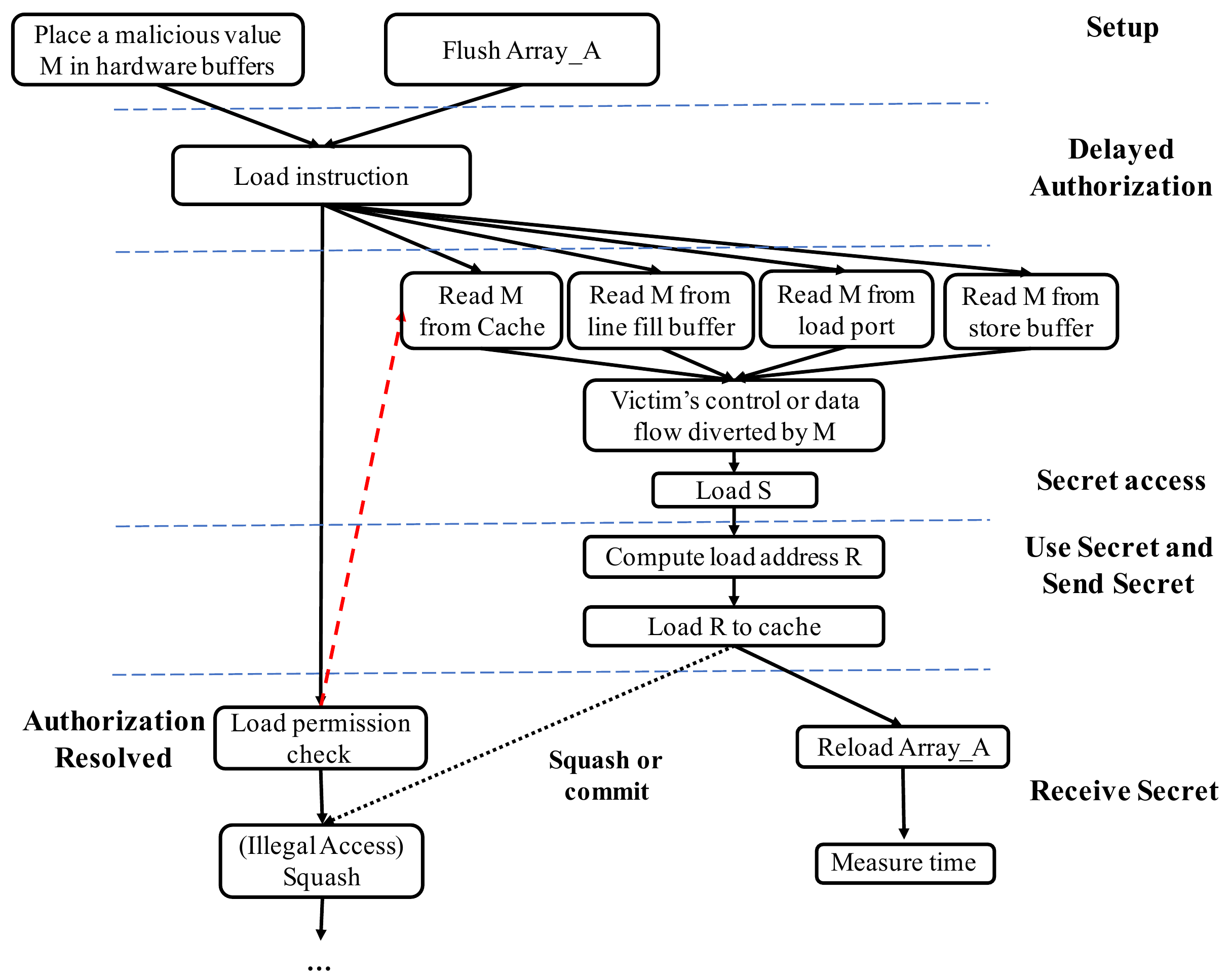}
    \caption{TSG model of Load Value Injection (LVI).} \label{fig:TSG:more-attack-lvi}
    \vspace{-10pt}
\end{figure}

The LVI attack injects the attacker-desired data to the victim's program. In this attack, the attacker attempts to leave the data in the memory buffers. A victim's faulting load speculatively reads from the buffer and unintentionally uses the attacker-controlled data for his execution. We model this attack in Figure \ref{fig:TSG:more-attack-lvi}. The missing security dependency is the red dashed arrow from the load fault handling to the access to the malicious data M.

A few of the entries in Table \ref{tab:TSG:AAnodes} have not been specifically described. Spectre v1.1 and Spectre v1.2 are like Spectre v1 and can be modeled by Figure \ref{fig:TSG:spectre} with a small modification. Instead of reading an out-of-bounds memory location, Spectre v1.1 writes an out-of-bounds memory location illegally. Spectre v1.2 tries to write to a read-only memory location.

Spectre RSB is like Spectre v2 (indirect branch). Hence, it can also be modeled by Figure \ref{fig:TSG:spectre}. Instead of waiting for the target address of an indirect branch instruction to be computed, Spectre RSB waits for the return address to be determined.


The last two entries in Table \ref{tab:TSG:AAnodes}, TAA and Cacheout, are TSX-based attacks. TSX uses transactional memory to enforce the atomic execution of a bundle of instructions called a transaction - either all the instructions are executed or none are executed. Hence, TSX can also be used to speculatively access a secret from the cache, store/load buffers or fill buffers.

\section{Benefits of our Model} \label{sec:application}

Our new attack graph model is useful in enabling us to:
\begin{itemize}

\item discover new attacks (Section \ref{sec:app:attack}),

\item model defense strategies and consistently explain why a specific defense works or does not work (Section \ref{sec:app:defense}),

\item enable tools to discover vulnerabilities and patch them (Section \ref{sec:app:tool}).

\end{itemize}

\subsection{Finding New Attacks}
\label{sec:app:attack}

Our attack graph can be generalized to model or find new attacks. We describe three ways: by finding new sources of secrets, new exploitable hardware features for delaying authorization, and new covert channels.

First, as we have already illustrated in Section \ref{sec:TSG}, the attack graphs can be extended to incorporate new sources of a secret. For instance, the micro-architectural data sampling attacks (RIDL \cite{van2019ridl}, Fallout \cite{canella2019fallout}, ZombieLoad \cite{schwarz2019zombieload}) use a faulting load to read secret data that is left in micro-architectural data buffers by previous memory accesses even from a different thread or process. They can be identified by analyzing the hardware implementation as the hardware designer should be able to find a set of datapaths that read data from different data buffers and forward the data to the faulting load. Each of these datapaths can be added as a new node in the attack graph (see Figure \ref{fig:TSG:meltdown}). 
Also, in the Meltdown variant1 \cite{meltdown3a} and LazyFP \cite{stecklina2018lazyfp} attacks, the unauthorized access to system registers or floating-point registers will cause an exception, and can be modeled with the nodes ``Read from Special Register'' or ``Read from FPU'' (see Figure \ref{fig:TSG:more-attack-register}). Other sources of secrets can also be identified to create new attacks.

Second, new hardware features can be exploited for delaying the authorization while allowing the execution to proceed. Examples include other hardware prediction mechanisms or delayed exception mechanisms. Identifying new authorization-related features can be achieved by analyzing processor pipeline squash signals. Each cause of a potential pipeline squash can be studied for its effect at the instruction (software) level. In the example of a conditional branch, the cause of the pipeline squash is due to the resolution of a conditional branch prediction. Subsequent load instructions after this conditional branch instruction can be the \emph{access} of a secret, followed by a covert \emph{send} through the cache covert channel, which gives rise to the Spectre v1 attack. 

In general, any decision-making operation that can cause speculative prediction and execution can trigger subsequent illegal accesses through a speculatively-executed load instruction (or privileged register read). This can be a ``software authorization'' node that triggers subsequent illegal accesses. Such decision-making actions can be expressed in any software language.

Furthermore, speculative execution is not the only source of transient instructions for illegally accessing secrets. Another example of transient instructions that may be aborted is TSX, for the atomic execution of a transaction, as also described earlier in the last two entries of Table III.

Third, our attack graph can also be extended to various different covert channels. For the most representative cache covert channel, we can generalize the cacheline as a resource whose state can be changed by the covert-sender or victim program, and this state change can be observed by the covert-receiver (attacker). To extend the analysis to different covert channels such as the memory bus covert channel, functional unit covert channel or branch target buffer (BTB) covert channel, we can also model the covert channel state and find the instructions that change this state and be detectable by a covert-receiver. This method can identify more sender-receiver pairs than the ``Load R to Cache'' and ``Reload Array\_A'' pair.

\textit{The key takeaway of this framework is that any new combination of these three dimensions of an attack gives a new attack.}

\subsection{Identifying Defense Strategies} \label{sec:app:defense}

\begin{figure}[t!]
    \centering
    \includegraphics[width=\linewidth]{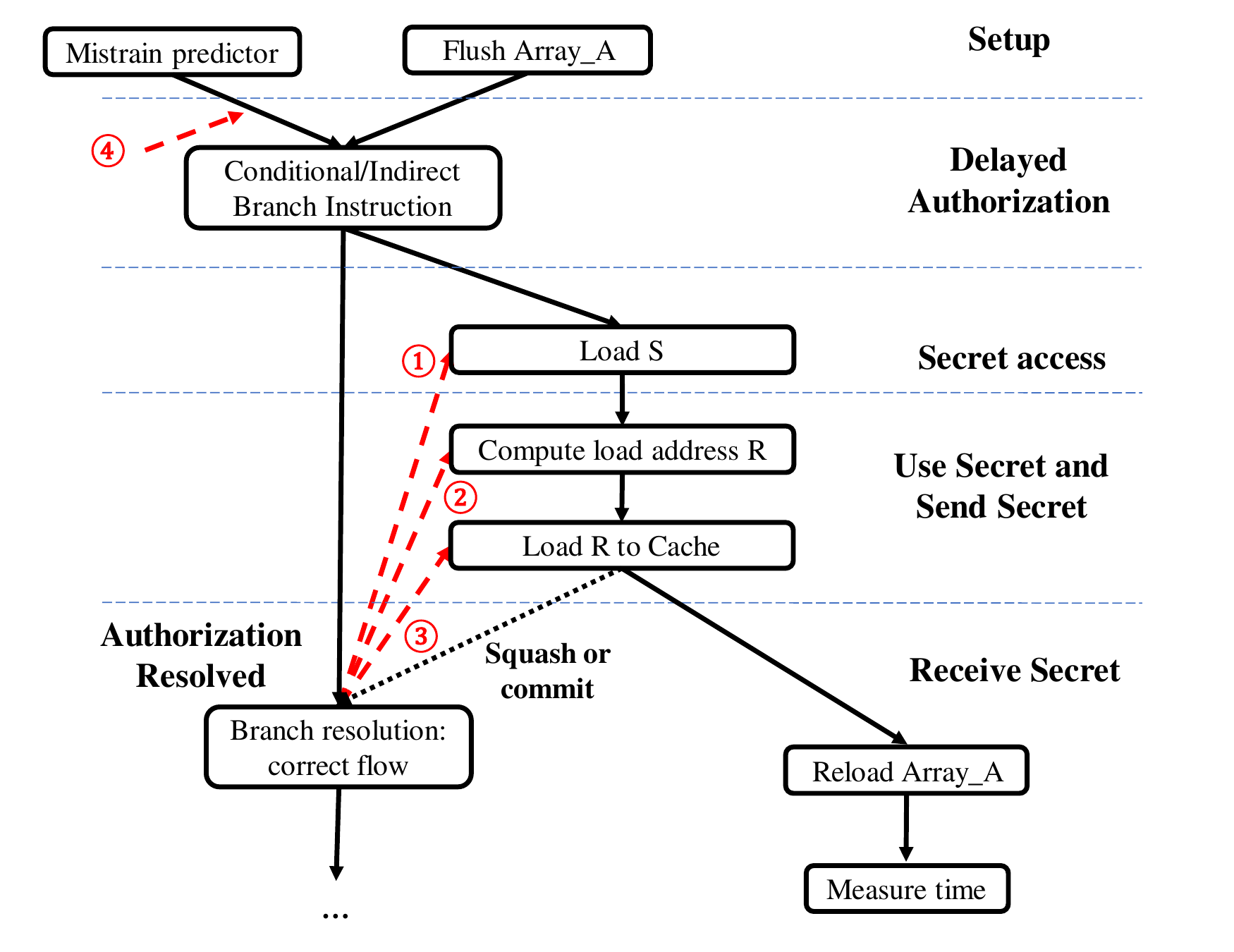}
    \caption{Four defense strategies against Spectre v1/v2 attacks: \ding{172} Add a security dependency between the ``branch resolution'' (authorization) for bounds checking, and the protected memory access, \ding{173} avoid the usage of speculative data, \ding{174} prevent loads whose address depends on unauthorized data from changing cache states, \ding{175} Clear predictor on context switch.}
    \label{fig:application:defense:spectre}
    \vspace{-10pt}
\end{figure}

A major application of our attack graph model is identifying potential defense strategies, as we illustrate by the red dashed arrows in Figure \ref{fig:application:defense:spectre} for the attacks triggered by branch instructions.
We illustrate potential defenses that essentially add security dependencies to the system to defeat the attacks. We also show that our defense strategies cover the recently proposed defenses in industry and academia to defeat speculative execution attacks.

\bheading{Strategy \ding{172}: Prevent Access before Authorization.} This prevents the illegal access of the secret, until the delayed authorization is resolved.

LFENCE is an industry defense used to serialize the instructions before and after it. Adding an LFENCE instruction before the speculative load adds a new security dependency between the ``Branch resolution'' (software authorization) and ``Load S'' (secret access), shown as \ding{172} in Figure \ref{fig:application:defense:spectre}. Context-sensitive fencing \cite{taram2019context} prevents the speculative access by inserting fences at the micro-operation level, e.g., between a conditional branch and a load to defeat the Spectre v1 attack. This is done in hardware, rather than in software. Secure automatic bounds checking (SABC) \cite{sabc:ojogbo2020secure} serialize the branch and the out-of-bounds access to mitigate the Spectre attack, by inserting arithmetic instructions with data dependencies between the branch and the access.

\bheading{Strategy \ding{173}: Prevent Data Usage before Authorization.} This prevents the use of the speculatively accessed secret, until the delayed authorization is resolved.

NDA \cite{weisse2019nda}, SpecShield \cite{barber2019specshield}, SpectreGuard \cite{fustos2019spectreguard} and ConTExT \cite{schwarz2020context} prevent forwarding the speculatively loaded data to the
following instructions so that the secret cannot be used, e.g., to compute the address $R$. SpectreGuard and ConTExT further provide the software interface for software developers to mark memory regions containing the secret as sensitive so the usage of non-sensitive data is allowed to reduce the performance overhead. Equivalently, this means adding a new security dependency between the ``Branch resolution'' (software authorization) and ``Compute Load Address R'' (data usage), shown as \ding{173} in Figure \ref{fig:application:defense:spectre}.

\bheading{Strategy \ding{174}: Prevent Send before Authorization.} This prevents the micro-architectural state changes of shared hardware resources that serve as the Send signal of the covert or side channel, until the delayed authorization is resolved. This defense strategy improves performance by adopting a looser security model where the secret is allowed to be accessed before authorization as long as it does not leak out.

This strategy adds the security dependency between ``Branch resolution'' and ``Load R to cache'' (cache state change), shown as \ding{174} in Figure \ref{fig:application:defense:spectre}. Different hardware implementations have been proposed under this strategy. STT \cite{yu2019speculative} and SpecShieldERP+ \cite{barber2019specshield} prevent loads whose address is based on speculative data. Conditional Speculation \cite{li2019conditional} and Efficient Invisible Speculative Execution \cite{sakalis2019efficient} both allow a speculative load that hits in the cache, because the cacheline state does not change on a hit, but delay speculative loads that encounter a miss. They further reduce the overhead by identifying trusted pages and predicting data, respectively. InvisiSpec \cite{yan2018invisispec} and SafeSpec \cite{khasawneh2018safespec} disallow speculative cache state modification but put the speculatively requested cache line in the shadow buffer. If the prediction is later found to be correct, InvisiSpec and SafeSpec reissue the memory access to fetch the proper cache lines. CleanupSpec \cite{saileshwar2019cleanupspec} allows speculative cache state modification but restores the cache state on a mis-speculation.


\bheading{Strategy \ding{175}: Clearing Predictions.} This strategy prevents the sharing of predictor states between different contexts.

For example, the industry solution from Intel, Indirect Branch Predictor Barrier (IBPB) \cite{ibpb}, prevents the code before the barrier from affecting the branch prediction after it by flushing the Branch Target Buffer (BTB). It introduces a new operation, i.e., ``flush predictor'', to the attack graph and adds a security dependency between ``flush predictor'' and the indicated branch instruction. Context-sensitive fencing\cite{taram2019context} also shows the feasibility of inserting micro-ops by hardware during a privilege change, to prevent a predictor being mistrained from a different context.

We show that our attack graph can model not only the defenses that work, but also the defenses that do not work. In Figure \ref{fig:TSG:meltdown}, we show that a security dependency can be added at four different places to defend against the Meltdown attack, shown as red dashed lines.  Defense strategies \ding{172}, \ding{173}, \ding{174} are similar to Figure \ref{fig:application:defense:spectre}. Typically, only one of these defense strategies is needed.

However, sometimes a defense is not sufficient, as we now illustrate with a hypothetical Meltdown attack coupled with an attacker induced cache hit for the secret, like the L1 Terminal Fault \cite{van2018foreshadow}. If the secret is already in the cache, the load instruction will fetch it from the cache rather than from the main memory. So while dependency \ding{172} can defend against the baseline Meltdown attack that speculatively loads a secret from main-memory in Figure \ref{fig:TSG:meltdown}, this is insufficient when \ding{172} can no longer prevent the secret access from the cache. In this case, an additional dependency \ding{175} in Figure \ref{fig:TSG:meltdown}, i.e. from ``Authorization'' to ``Read S from cache'', has to be jointly added with \ding{172} to provide a valid defense. Hence, it is important to put security dependencies in the correct places, otherwise we get a false sense of security, especially when micro-architectural performance optimizations (like load from cache on a hit) can bypass an insecure security dependency like \ding{172}. In fact, there has to be a security dependency arrow between the ``authorization resolved'' node to every node that can be a source of the secret in Figure \ref{fig:TSG:meltdown}, such as load ports, line fill buffers and store/load buffers. The number of such ports and buffers suggests that the defense strategy ``Prevent Data Usage before Authorization'' may be a solution that is not only more efficient but also more secure.

\subsection{Tools for Constructing Attack Graphs}
\label{sec:app:tool}

A tool can be designed to construct attack graphs and find the missing security dependencies. To achieve this, the memory locations (with secret or sensitive data and code) to be protected should be identified. OS and hypervisor data and code must automatically be protected from less-privileged software. For user-level data and code, the most secure way is for the user to initially specify what data and code should be protected as in  \cite{fustos2019spectreguard, schwarz2020context}. Then the tool can trace all direct and indirect accesses to these protected data and code as potential secret accesses.

\begin{figure}[t]
    \centering
    \includegraphics[width=\linewidth]{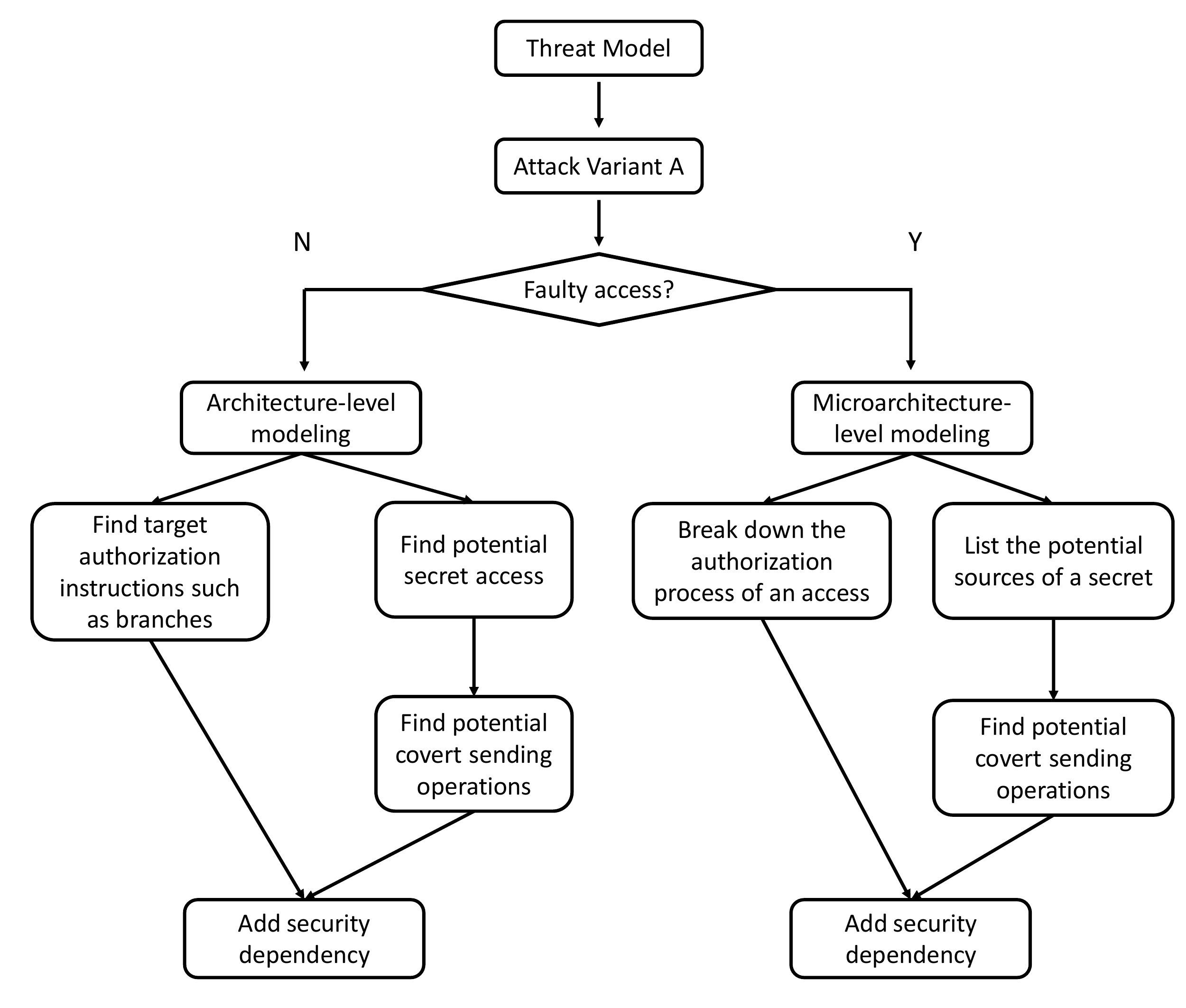}
    \caption{The flow chart to generate the attack graph for different types of speculative execution attacks. This also shows how the vulnerability can be plugged by adding a security dependency.}
    \label{fig:app:graphgen}
    \vspace{-10pt}
\end{figure}

Then, the tool needs to identify attack nodes as we introduced in Section \ref{sec:TSG}. By providing a threat model that specifies the range of attacks that are to be defeated, the tool can recognize the authorization operation such as a prior conditional or indirect branch instruction (software authorization), or a load or store instruction (hardware privilege check or address disambiguation check). A flow chart to generate the attack graph for speculative execution attacks is shown in Figure \ref{fig:app:graphgen}.

For the control-flow misprediction attacks triggered by a conditional or indirect branch instruction (the left side of Figure \ref{fig:app:graphgen}), we propose a major simplification where these misprediction-based attacks can be modeled at the instruction level where the nodes are just instructions, and the edges are control flow and data-flow dependencies between instructions. This means that the tool just needs to look for subsequent memory loads or special register access instructions after branch (conditional or indirect) nodes as the secret access.

For the faulty memory/register access attacks where the authorization and the secret access are done in the same instruction, the tool needs to break down such instructions into their micro-architectural level in the attack graph model, as shown in Figure \ref{fig:TSG:meltdown}. These instructions to be broken down are the memory load instructions, the Read Privileged Register instructions and the instructions that Read or Write Floating-point or SIMD registers.

After identifying the secret access and the potential covert sending operations, the tool can automatically generate edges by looking for existing dependencies, e.g., data dependencies, fences and address dependencies. Missing security dependencies (races) between the authorization and the secret access instruction and its subsequent chain of data-dependent instructions, which can be executed in the speculative execution window, can be found by automatically searching the graph. Such a missing security dependency between an authorization node and a secret access node in a program shows a vulnerability that can lead to a potential attack. These vulnerabilities can be flagged by the tool. The tool can also proactively insert a security dependency, e.g., a lightweight fence, to prevent attacks. 

The main challenges for extending the above methodology to a fully-functional tool appear to be the initial identification of secrets in user code, and modelling at the intra-instruction level which is different for each  micro-architecture implementation. The former can leverage the use of other tools, and we have shown that the latter is only necessary for a few instructions where the authorization and access are in the same instruction, e.g., load instructions.

%
%



\section{Insights and Takeaways} \label{sec:insights}

Based on our new models of speculative execution attacks, our major new insights are:

\begin{enumerate}

\item The root cause of speculative attacks succeeding is a missing edge in the attack graph between the authorization operation and the secret access operation. 

\item We define the term ``security dependency'', and equate this to a missing edge in an attack graph that enforces the correct ordering of the authorization node before subsequent operation nodes.

\item These security dependencies can give us ``defense strategies''. Each defense strategy can be implemented through many different architectural mechanisms. More importantly, the currently proposed hardware and software defenses all fall under one of our defense strategies.

\item Falling under one of our defense strategies also explains why the defense works. The defense is equivalent to implementing a missing security dependency and thus preventing a critical attack step from succeeding. This is the first time a reason for the success of a defense can be systematically given.

\item A security dependency can sometimes be ``relaxed'' to reduce the performance overhead (e.g., allow accessing the secret but prevent leaking the secret) for security-performance tradeoffs. This is illustrated in Section \ref{sec:app:defense} by our defense strategies \ding{173} prevent data usage before authorization and \ding{174} prevent send before authorization.

\item Attacks that look similar, e.g., the Spectre-type and Meltdown-type attacks are actually different, in the sense that the Meltdown-type attacks have authorization and access in the same instruction, while the Spectre-type attacks do not. This results in the Meltdown-type attacks having to be investigated through intra-instruction operations, while Spectre-type attacks only need to consider inter-instruction operations. This makes attack graph and tool construction simpler.

\end{enumerate}

Our new attack graph is useful in finding new attacks,
 identifying new defense strategies and systematically explaining why a defense works or not,
 and facilitating the design of  tools to discover vulnerabilities and patch them.
\section{Related Work}
\label{sec:related}

Speculative execution attacks have been defined, e.g., in \cite{kocher2018spectre, lipp2018meltdown, stecklina2018lazyfp, kiriansky2018speculative, koruyeh2018spectre, spectreSSB, van2018foreshadow}. To model speculative execution attacks, Disselkoen \etal \cite{disselkoencode} proposed a memory model based on pomsets to allow speculative evaluation. However, this model only covers speculative secret access, but does not show how the secret is sent through a cache covert channel. Canella \etal \cite{canella2018systematic} summarized and evaluated speculative execution attacks and some defenses. However, their work does not provide a systematic model for attacks and defenses that provides insights on designing and evaluating new secure defenses. On the formal modeling side, Guarnieri \etal \cite{guarnieri2018spectector} proposed the speculative non-interference property to verify that a program behaves the same with and without speculation. Cheang \etal \cite{cheang2019formal} proposed trace property-dependent observational determinism (TPOD) to verify that two traces of execution are not distinguishable. These formal methods cannot reason about the defenses, while we can show why defenses work and which ones will not work.

Previous work has been proposed to evaluate caches' resilience against (non-speculative) side-channel attacks. He \etal \cite{he2017secure} proposed the probabilistic information flow graph (PIFG) to model the cache, attacker and victim simultaneously. Zhang \etal \cite{zhang2014new} quantified information leakage via mutual information. Demme \etal \cite{demme2012side} proposed the Side-channel Vulnerability Factor (SVF) and Zhang \etal \cite{zhang2013side} refined it as the Cache Side-channel Vulnerability (CSV) for access-based attacks. However, none of these past work on cache side-channel attacks covers speculative execution attacks.

\section{Conclusions} \label{sec:conclu}

Information leakage due to speculative execution has been a serious and unsolved problem. In this paper, we provide new attack graph models for speculative execution attacks. We identify the common characteristics of speculative attacks as illegal access of secrets during speculative execution and covert channel exfiltration of this secret information, and break these down further into 5-6 critical attack steps.

We propose the attack graph as a topological sort graph (TSG), where the critical attack steps can be identified. Fundamentally, the speculation vulnerability is due to a race condition, shown as a missing edge in a TSG, between authorization and secret access nodes. In a looser security but higher performance scenario, this missing edge can be between the authorization node and the nodes that \textit{use} or \textit{send} out the ``not-yet-authorized'' data. We are the first to define the concept of a \textit{security dependency}, which enforces the proper ordering of \textit{authorization before access}, or \textit{authorization before use}, or \textit{authorization before send} operations. Security dependencies prevent race conditions that lead to security breaches.

To show the effectiveness of our proposed models, we generate attack graphs for the Spectre and Meltdown attacks, and then generalize them to all known attack variants. From our attack graphs, we show how to generate new attacks, how to derive new defense strategies and why they work. In fact, all proposed solutions from both industry and academia fall under one of our defense strategies. We have also shown how to design a tool that can help construct attack graphs and uncover vulnerabilities in the software-hardware system. 

We have provided a generalizable framework to model and analyze the speculative execution attacks, and hope this helps advance more secure micro-architecture defenses and designs.


\section*{Acknowledgements}
This work is supported in part by NSF SaTC \#1814190, SRC Hardware Security Task 2844.002 and a Qualcomm Faculty Award for Prof. Lee. We thank the anonymous reviewers for their insightful comments and feedback.

\begin{appendices}
\section{Proof of Theorem 1} \label{sec:appendix}

\setlength{\abovedisplayskip}{0pt}%
\setlength{\belowdisplayskip}{0pt}%
\setlength{\abovedisplayshortskip}{0pt}%
\setlength{\belowdisplayshortskip}{0pt}%
\setlength{\jot}{0pt}

\textit{\textbf{Proof: Necessity (<=).}} It can be proved by contradiction. Assume there is not a path from $u$ to $v$. Let
\begin{align}
S&=[s_1, s_2...s_k, v, s_{k+1} ... s_n]\\
&=[S_v, v, s_{k+1} ... s_n]
\end{align}

\noindent be a \textbf{valid} ordering sequence, where $S_v=[s_1, s_2...s_k]$ represents the vertices before $v$ in $S$. Split $S_v$ into two subsequences with the same order in $S$
\begin{align}
S_1 = [s_1^1, s_1^2...s_1^{k_1}]\\
S_2 = [s_2^1, s_2^2...s_2^{k_2}]
\end{align}

where $S_1$ contains vertices that have a path to $v$, $S_2$ contains vertices that do not have a path to $v$. By assumption, $u$ does not have a path to $v$, therefore $u\in S_2$. Note that $k_1+k_2=k$. Construct another sequence
\begin{align*}
S'&=[S_1, v, S_2, s_{k+1} ... s_n]\\
&=[s_1^1, s_1^2...s_1^{k_1}, v, s_2^1, s_2^2...s_2^{k_2}, s_{k+1} ... s_n]
\end{align*}

We claim that ordering $S'$ is also valid:
For any $s_1^{k_i} \in S_1$ and $s_2^{k_j} \in S_2$, there is not an edge ($s_2^{k_j}$, $s_1^{k_i}$) in the graph. Otherwise, [$s_2^{k_j}\to s_1^{k_i}\to v$] is a path, contradicting the definition of $S_2$. For the same reason, there is not an edge ($s_2^{k_j}$, $v$) in the graph. We categorize any edge ($z$, $x$) into 3 cases, i.e. $x \in S_1$, $x=v$ and $x\in S_2$. We show $z$ comes before $x$ in $S'$ in all cases:\\
\ding{172} for any edge ($z$, $v$) in the graph, $z$ can only be in $S_1$, and thus $z$ is before $v$ in $S'$.\\
\ding{173} For any edge ($z$, $s_1^{k_i}$) in the graph, $z$ can only be in $(s_1^{k_1},...s_1^{k_{i-1}})$, and thus $z$ is also before $s_1^{k_i}$ in $S'$.\\
\ding{174} Since $S_2$ are moved backward and $s_{k+1}...s_{n}$ are kept in the same position from $S$ to $S'$, for any edge ($z$, $s_2^{k_j}$), $z$ is before the $s_2^{k_j}$.\\
From \ding{172},\ding{173} and \ding{174}, the ordering $S'$ is valid. Meanwhile, $v$ is before $u\in S_2$ in $S'$, contradicting to $u$ is before $v$ in all valid orderings. Therefore, the assumption is not true and there must be a path connecting $u$ and $v$. \QEDB

\textit{\textbf{Sufficiency (=>).}}
The sufficiency is relatively obvious. Without loss of generality, assume there exists a directed path from $u$ to $v$, i.e. $P$=$(u, w_1, ... w_k, v)$. Then for any valid ordering $S$, $u$ is before $w_1$, $w_1$ is before $w_2$, ... ,$w_k$ is before $v$. Therefore $u$ is before $v$ in any valid ordering. \QEDB

\end{appendices}

\bibliographystyle{IEEEtranS}
\bibliography{main}

\end{document}